\def\papertitle{FM Synthesizer Audio-Parameter Shared Embeddings}
\def\paperauthorA{David Braun}
\def\paperauthorB{Adam Finkelstein}

\documentclass[twoside,a4paper]{article}
\usepackage{etoolbox}

\usepackage[print]{dafx26v3}

\usepackage{amsmath,amssymb,amsfonts,amsthm}
\usepackage{siunitx}
\usepackage{euscript}
\usepackage[T1]{fontenc}
\usepackage[utf8]{inputenc}
\usepackage{ifpdf}
\usepackage[english]{babel}
\usepackage{caption}
\usepackage{subfig} 
\usepackage{color}
\usepackage{booktabs}
\usepackage{lipsum}

\input glyphtounicode
\ninept

\newcounter{numauth}
\newcounter{listcnt}
\newcommand\authcnt[1]{\ifdefined#1 \stepcounter{numauth} \fi}

\newcommand\addauth[1]{
\ifdefined#1
\stepcounter{listcnt}
\ifnum \value{listcnt}<\value{numauth}
\appto\authorslist{, #1}
\else
\appto\authorslist{~and~#1}
\fi
\fi}
\authcnt{\paperauthorB}
\authcnt{\paperauthorC}
\authcnt{\paperauthorD}
\authcnt{\paperauthorE}
\authcnt{\paperauthorF}
\authcnt{\paperauthorG}
\authcnt{\paperauthorH}
\authcnt{\paperauthorI}
\authcnt{\paperauthorJ}
\def\authorslist{\paperauthorA}
\addauth{\paperauthorB}
\addauth{\paperauthorC}
\addauth{\paperauthorD}
\addauth{\paperauthorE}
\addauth{\paperauthorF}
\addauth{\paperauthorG}
\addauth{\paperauthorH}
\addauth{\paperauthorI}
\addauth{\paperauthorJ}

\usepackage{times}

\newif\ifpdf
\ifx\pdfoutput\relax
\else
   \ifcase\pdfoutput
      \pdffalse
   \else
      \pdftrue
   \fi
\fi

\ifpdf 
  \usepackage[pdftex,
    pdftitle={\papertitle},
    pdfauthor={\authorslist},
    pdfsubject={Proceedings of the 29th International Conference on Digital Audio Effects (DAFx26)},
    colorlinks=false, 
    bookmarksnumbered, 
    pdfstartview=XYZ 
  ]{hyperref}
  \usepackage[pdftex]{graphicx}
\else 
  \usepackage[dvips]{epsfig,graphicx}
  \usepackage[dvips,
    pdftitle={\papertitle},
    pdfauthor={\authorslist},
    pdfsubject={Proceedings of the 29th International Conference on Digital Audio Effects (DAFx26)},
    colorlinks=false, 
    bookmarksnumbered, 
    pdfstartview=XYZ 
  ]{hyperref}
\fi
\usepackage[hypcap=true]{caption}
\title{\papertitle}

\affiliation
{\paperauthorA\ and \paperauthorB}
{\href{https://cs.princeton.edu}{Dept. of Computer Science} \\ Princeton University \\ Princeton, NJ, USA\\
{\tt \href{mailto:db1224@princeton.edu}{db1224@princeton.edu}}
}

\begin{document}
\ifpdf 
  \DeclareGraphicsExtensions{.png,.jpg,.pdf}
\else  
  \DeclareGraphicsExtensions{.eps}
\fi


\maketitle

\begin{abstract}
Given a target sound, finding the synthesizer preset that best reproduces it remains a core problem in sound design.
Existing methods treat synthesis parameters as flat vectors, discarding the signal routing and parameter interactions that produce audio.
We make two contributions.
First, to learn a representation of parameters including their signal routing, we design a graph neural network whose message passing structure imitates FM signal processing.
Second, we adapt the multimodal objective from SLAP to learn joint embeddings of audio and FM synthesizer parameters, enabling preset retrieval from a gallery.
We focus on the Yamaha DX7, where six identical sinusoid operators interact according to one of 32 routing topologies.
Our graph encoder's message passing weights are shared across all nodes and layers, enabling processing of arbitrary topologies of any size.
When every topology is seen during training, the DX7-GNN and two baselines achieve strong audio-to-preset retrieval.
When some topologies are held out for testing, the DX7-GNN substantially outperforms both baselines despite having the fewest parameters.
Our ablations further support the claim that imitating FM signal flow in a parameter encoder improves generalization to unseen topologies.

\end{abstract}

\section{Introduction}
\label{sec:intro}

Sound designers routinely search for synthesizer presets by ear, a slow process that scales poorly across large preset libraries.
To aid in this task, automatic synthesizer programming~\cite{shier_synthesizer_2021} methods rely on supervised regression~\cite{yee-king_automatic_2018}, gradient-based optimization~\cite{yang_white_2023}, evolutionary search~\cite{masuda_quality-diversity_2023}, or reinforcement learning~\cite{shin_synthrl_2025}.
Joint audio-text embeddings like CLAP~\cite{elizalde_clap_2023,wu_large-scale_2023} support synthesizer sound search and optimization~\cite{brade_synthscribe_2024,cherep_creative_2024}, but text descriptions are coarse, subjective, and require labor-intensive human labels.

A joint embedding of audio and synthesis parameters would combine the retrieval and optimization affordances of CLAP with the precision of parameter-level representations.
However, few methods learn reusable parameter representations~\cite{combes_neural_2025,le_vaillant_latent_2024,hayes_audio_2025}, and most ignore the computational structure that defines how parameters interact to produce sound.
Frequency Modulation (FM) synthesis~\cite{chowning_synthesis_1973} illustrates why structure matters.
The topology of operator connections shapes the timbre, and no single FM topology is optimal across all desired sounds~\cite{ye_nas-fm_2023}.
Existing methods encode the routing as a \textit{categorical} label rather than representing the topology itself, so they cannot generalize to new routings.

We address these limitations by learning a shared embedding space for audio and FM synthesizer presets with a topology-aware parameter encoder.
We focus on the Yamaha DX7, a commercially popular synthesizer widely used in FM synthesis research.
Its six operators (parameterized sine oscillators) are arranged according to one of 32 predefined routing ``algorithms'' that specify modulation and carrier relationships.
This structure naturally corresponds to a graph neural network (GNN) where operators are nodes and modulation connections are directed edges.
Messages flow along these edges like modulation signals in the synthesizer: each node's output is gated by its operator's output level, incoming messages are summed, and the aggregate features modulate the node's base features via FiLM conditioning.
All message passing weights are shared across nodes and layers, so the model can encode topologies never seen during training.
Most research on FM synthesis excludes feedback due to its complexity~\cite{ye_nas-fm_2023,caspe_ddx7_2022,uzrad_diffmoog_2024,turian_one_2021};
our graph structure represents feedback as ordinary messages attenuated with learned edge weights.

We make two primary contributions:
\begin{enumerate}
\item We adapt the multimodal objective from SLAP~\cite{guinot_slap_2025} to learn joint embeddings of audio and FM synthesizer parameters. We call this framework FM-SynAPSE: \textbf{Syn}thesizer \textbf{A}udio-\textbf{P}arameter \textbf{S}hared \textbf{E}mbeddings. Given target audio, our system retrieves the closest preset from a gallery, instead of predicting or optimizing parameters.
\item We introduce DX7-GNN, a graph neural network parameter encoder. Its message passing architecture imitates FM signal processing and generalizes to synthesizer topologies never seen during training, a capability flat parameter encoders lack.
\end{enumerate}

A controlled comparison with Transformer and Highway Network baselines confirms the value of our structural inductive bias.
When all 32 DX7 algorithms appear during training, all three encoders achieve strong retrieval performance.
Using 16 diverse algorithms for training and eight for testing, the DX7-GNN substantially outperforms both baselines despite having the fewest parameters.
Although trained only on FM synthesis, our audio encoder reaches 69\% triplet agreement on a timbre similarity benchmark, approaching LAION-CLAP's 72\%.
To support reproducibility, we release our codebase, model weights, and an interactive retrieval website.\footnote{\url{https://github.com/DBraun/SynAPSE}}

\section{Related Work}
\label{sec:related_work}

\textbf{Automatic synthesizer programming.}
Estimating synthesizer parameters from target audio has been approached through supervised regression~\cite{yee-king_automatic_2018,vaillant_improving_2021,bruford_synthesizer_2024,barkan_inversynth_2023}, discretization with label smoothing~\cite{chen_sound2synth_2022}, reinforcement learning~\cite{shin_synthrl_2025}, and generative modeling of parameter distributions~\cite{hayes_audio_2025}.
Optimization approaches include differentiable rendering with gradient search~\cite{yang_white_2023} and quality-diversity search~\cite{masuda_quality-diversity_2023}.
These methods predict or optimize parameters for a fixed synthesizer architecture.

For FM synthesis, flexible topologies add another challenge. NAS-FM~\cite{ye_nas-fm_2023} searches jointly over FM topologies and frequency ratios with evolutionary methods, finding that no single topology is optimal across instruments.
DiffMoog~\cite{uzrad_diffmoog_2024} implements a differentiable modular synthesizer but reports that its FM modulation chains fail to converge under gradient-based optimization.
Le Vaillant et al.~\cite{le_vaillant_latent_2024} learn DX7 preset encodings for latent space interpolation but encode the algorithm as a categorical parameter.
Hayes et al.~\cite{hayes_audio_2025} propose \texttt{Param2Tok} to encode parameters and discover permutation symmetries among interchangeable oscillators and filters.
However, the signal routing is still a one-hot parameter, so the method cannot encode a preset for an unseen topology.

\textbf{Learned parameter representations.}
Instead of directly regressing audio to parameters, recent work learns reusable parameter encoders.
Synth-Proxy~\cite{combes_neural_2025} trains a preset encoder to approximate embeddings from a frozen audio model, comparing Highway Network, Transformer, and other architectures across three synthesizers.
Their encoders treat parameters as flat vectors, and each model is trained on a fixed synthesizer architecture.
We adopt their Highway Network as a baseline.
InverSynth~II~\cite{barkan_inversynth_2023} learns a differentiable synthesizer-proxy that maps flat parameter vectors to spectrograms, enabling audio-based and parameter-based losses for sound matching.
Its proxy also treats parameters as a flat concatenation of one-hot and scalar values.

\textbf{Learned audio representations.}
Other work uses diverse synthesizer presets to learn timbre-relevant audio representations without explicitly encoding the parameters.
Cherep and Singh~\cite{cherep_contrastive_2025} render slightly perturbed copies of each preset to generate positive pairs for contrastive learning.
Le Vaillant and Molle~\cite{vaillant_contrastive_2026} learn contrastive timbre representations of sampler instruments and rendered synthesizer presets, retrieving instruments from sounds and mixtures by audio-to-audio matching.
Their retrieval gallery holds an embedding of each preset's rendered audio, whereas our parameter encoder embeds presets directly from their parameters.

\textbf{Multimodal embeddings.}
CLAP~\cite{elizalde_clap_2023,wu_large-scale_2023} learns joint audio-text embeddings via contrastive learning.
SLAP~\cite{guinot_slap_2025} replaces the contrastive objective with a non-contrastive, BYOL-style one~\cite{grill_bootstrap_2020} and outperforms CLAP on retrieval and downstream probing.
The non-contrastive objective also unlocks gradient accumulation, enabling large training batch sizes.

\section{Method}
\label{sec:method}

FM-SynAPSE learns joint embeddings of FM synthesizer parameters and audio.
Its parameter encoder, DX7-GNN, is built on the observation that FM synthesis has the structure of a message passing neural network.

\subsection{Problem Formulation}
\label{sec:problem}

Let $x_A \in \mathbb{R}^{T}$ denote a single-channel audio waveform of $T$ samples and $x_P$ denote a DX7 preset.
We learn an audio encoder and a parameter encoder that map $x_A$ and $x_P$ into a shared $d$-dimensional embedding space.
At inference time, the audio query embedding ($q_A$) is compared via cosine similarity against the parameter embeddings ($q_P$) of all presets in a gallery, and the nearest preset is returned.

\subsection{Training Objective: SLAP}
\label{sec:slap}

\begin{figure}[!t]
\centering
\includegraphics[width=0.9\columnwidth]{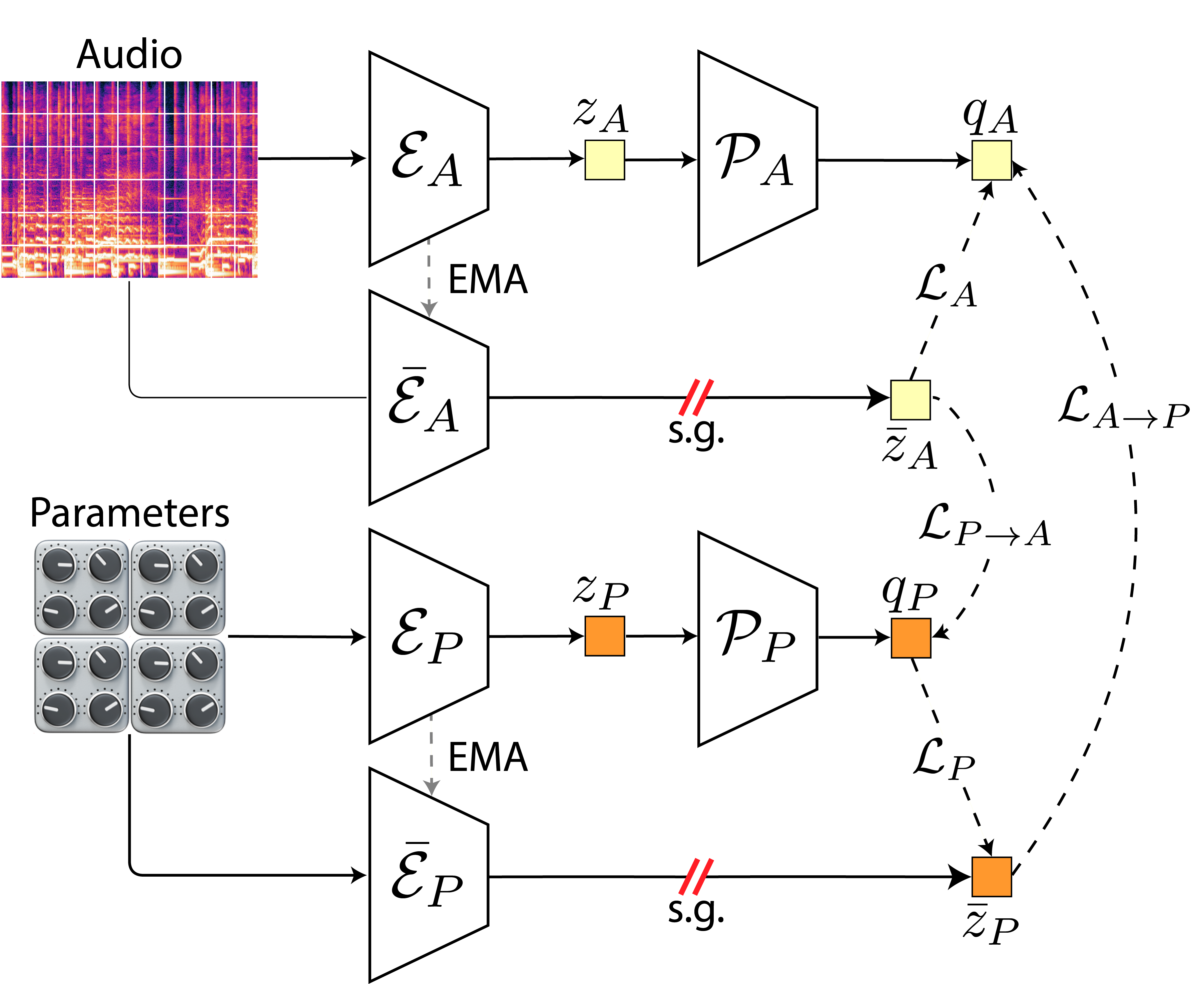}
\caption{
SLAP architecture for FM-SynAPSE: Online encoders ($\mathcal{E}_A$, $\mathcal{E}_P$) receive gradients;
Target encoders ($\bar{\mathcal{E}}_A$, $\bar{\mathcal{E}}_P$) are updated via exponential moving average (EMA).
Predictor networks ($\mathcal{P}_A, \mathcal{P}_P$) map online projections to predictions of target projections.
Four cosine similarity losses combine intermodal ($\mathcal{L}_{A \to P}$, $\mathcal{L}_{P \to A}$) and intramodal ($\mathcal{L}_A$, $\mathcal{L}_P$) alignment.
Dotted lines indicate how loss gradients backpropagate towards predictions, not target projections.
}
\label{fig:slap_architecture}
\end{figure}

We rely on the joint embedding objective introduced in SLAP~\cite{guinot_slap_2025}, where the language modality is replaced by the parameters of a DX7 preset (Figure~\ref{fig:slap_architecture}).
Each modality $M \in \{A, P\}$ has an online branch: the encoder $\mathcal{E}_M$ maps its input to a projection $z_M \in \mathbb{R}^{d}$, and a predictor $\mathcal{P}_M$ maps $z_M$ to a query $q_M \in \mathbb{R}^{d}$.
Inside $\mathcal{E}_M$, a backbone first produces an intermediate latent $y_M$ that a projector MLP maps to $z_M$.
A separate target branch $\bar{\mathcal{E}}_M$, identical to $\mathcal{E}_M$ but with no predictor, produces stop-gradient target projections $\bar{z}_M$.
Writing $\theta_M$ for the parameters of the online encoder and projector and $\bar{\theta}_M$ for the target's, the target is not trained by gradient descent but updated as an exponential moving average of the online branch:
\begin{equation}
\bar{\theta}_M \gets \tau\,\bar{\theta}_M + (1 - \tau)\,\theta_M, \qquad \tau = 0.98 .
\end{equation}
Together with the predictor, this asymmetry prevents representational collapse without negative pairs~\cite{grill_bootstrap_2020}.

Each loss term is a cosine distance $d_{\cos}(u, v) = 1 - \frac{u \cdot v}{\|u\|\,\|v\|}$.
The \emph{intermodal} losses align each modality's query with the other modality's stop-gradient target projection,
\begin{equation}
\mathcal{L}_{A \to P} = d_{\cos}(q_A, \bar{z}_P), \qquad
\mathcal{L}_{P \to A} = d_{\cos}(q_P, \bar{z}_A),
\end{equation}
and the \emph{intramodal} losses apply the same distance within each modality:
\begin{equation}
\mathcal{L}_{A} = d_{\cos}(q_A, \bar{z}_A), \qquad
\mathcal{L}_{P} = d_{\cos}(q_P, \bar{z}_P) .
\end{equation}
The total objective combines them with a weight $\lambda \in [0, 1]$:
\begin{equation}
\mathcal{L} = \lambda \big(\mathcal{L}_{A \to P} + \mathcal{L}_{P \to A}\big) + (1 - \lambda) \big(\mathcal{L}_{A} + \mathcal{L}_{P}\big) .
\end{equation}
We set $\lambda = 0.5$ following the ablations in~\cite{guinot_slap_2025}.
Three SLAP hyperparameters were adjusted during a hyperparameter search and carried over to all baselines: projector/predictor output dimension $d$ (512 $\to$ 384), EMA rate $\tau$ (0.95 $\to$ 0.98), and dropout rates (projector 15\%, predictor 35\%).
The heads otherwise follow SLAP.
Each is a two-layer ReLU MLP, and only the predictor uses a 4096-dimensional hidden layer and batch normalization.

\subsection{DX7 Setup}
\label{sec:dx7_setup}

Each DX7 operator is a sinusoid generator whose output is scaled by its output level, producing a signal in $[-1, 1]$.
Modulator operators pass their scaled output to shift the instantaneous phase of destination operators, creating complex timbres~\cite{chowning_synthesis_1973}.
An operator with output level zero produces no signal and therefore no modulation.
Carrier operators (those with no outgoing modulation connections) are summed to produce the final audio.
Each algorithm also has one single-sample delay feedback connection, controlled by a global feedback parameter.
In 30 algorithms, feedback is a self-loop (an operator modulating its own phase).
In Algorithms 4 and 6, feedback connects different operators.
Across the 32 algorithms, the number of modulations including feedback ranges from one (Algorithm 32, purely additive with only a feedback self-loop) to six (Algorithms 16--18, single-carrier with deep modulation chains), and carrier count ranges from one to six.
All algorithms appear in~\cite{ye_nas-fm_2023}.

\subsection{Graph-Based Parameter Encoder}
\label{sec:gnn}

\begin{figure}[t]
\centering
\includegraphics[width=0.49\columnwidth]{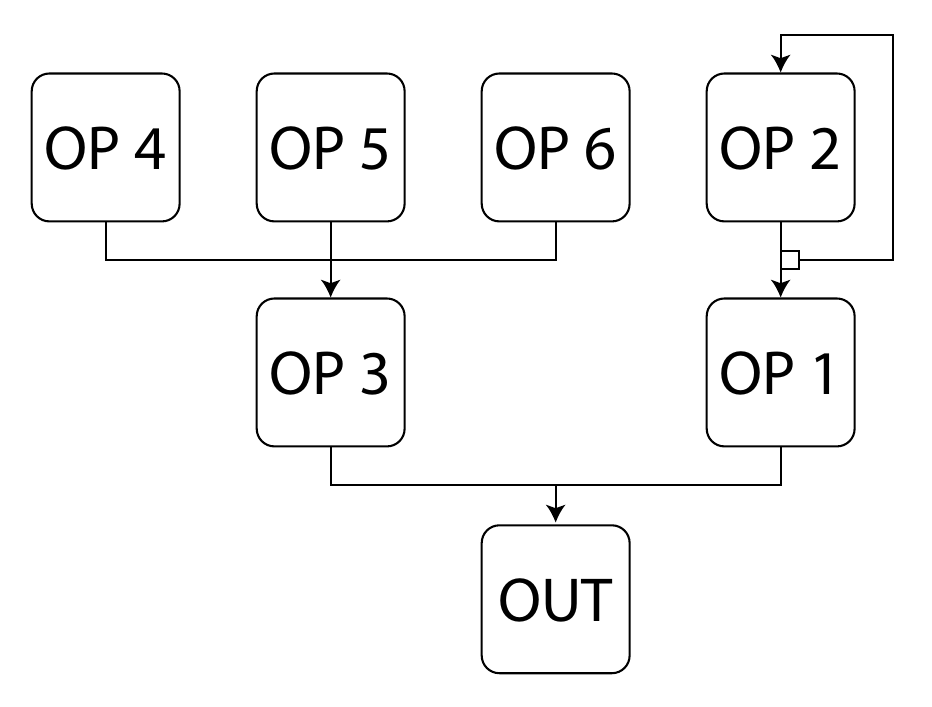}
\hfill
\includegraphics[width=0.49\columnwidth]{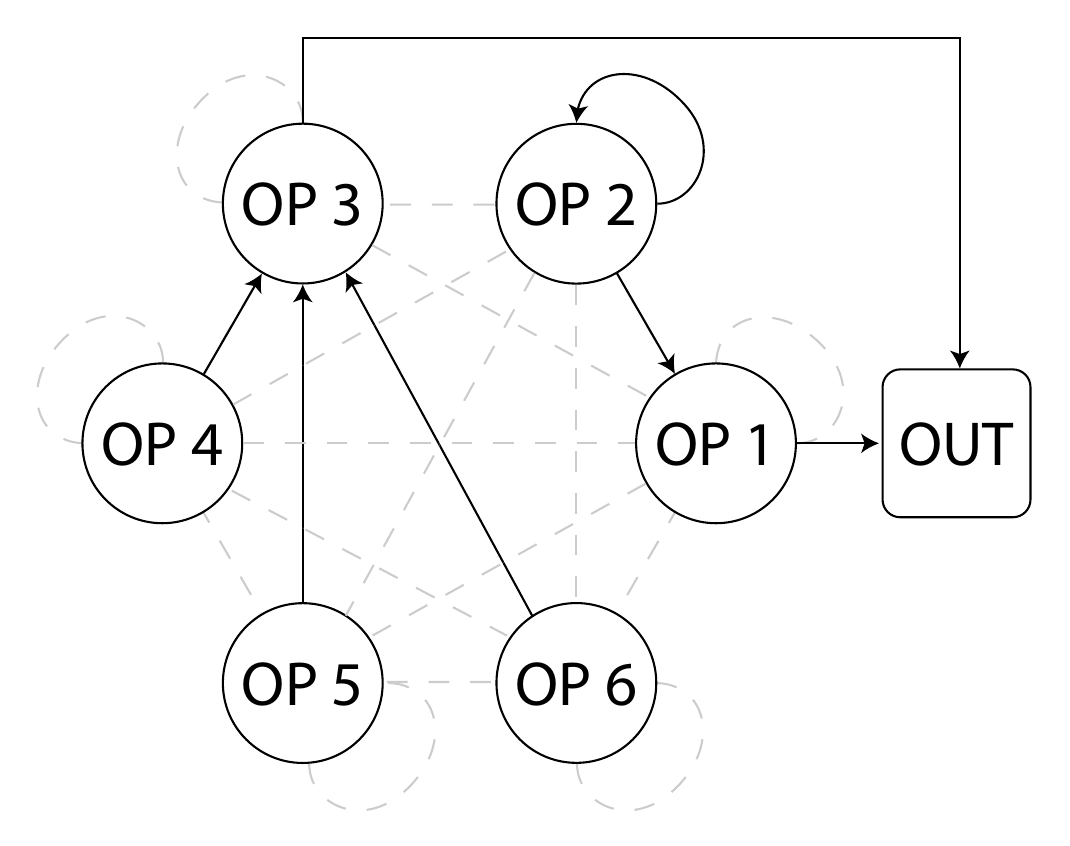}
\caption{Two views of Algorithm 12: (Left) Traditional DSP. (Right) A message passing graph.
Inactive edges are dotted lines.
}
\label{fig:fm_graphs}
\end{figure}

For each DX7 preset, we construct a directed graph $G = (V, E)$ determined by the selected algorithm (1--32): the six operators are the nodes $V = \{v_1, \dots, v_6\}$, and the algorithm's modulation connections and feedback loop are the edges $E$.
Figure~\ref{fig:fm_graphs} illustrates two views of Algorithm 12.

\subsubsection{Parameter Processing}
A DX7 preset $x_P$ consists of 145 parameters stored as a flat vector: 123 continuous floats in $[0, 1]$ and 22 integers.
These divide into 19 global parameters (feedback level, LFO settings, pitch envelope, algorithm selection) and 21 per-operator parameters (amplitude envelope, frequency, keyboard scaling) for each of six operators.
Rather than pool all six operators into a single vector, DX7-GNN gives each operator node its own feature vector.

Four parameters receive special handling instead of entering the per-node feature vector.
Three of them are global: the algorithm parameter instead determines the graph structure, transpose is fixed during rendering, and the feedback level enters the encoder \emph{only} through the feedback edge weight in message passing.
Any benefit from feedback must therefore arise from message passing rather than from a per-node feature.
The fourth is per-operator: the output level (six values, one per operator) instead scales the node features during message passing.
Excluding these leaves 16 of the 19 global parameters and 20 of the 21 per-operator parameters.
One-hot encoding expands categoricals: LFO waveform from one to six dimensions, left/right scaling curves from one to four each.
After encoding, the 16 global parameters become 21 dimensions and the 20 per-operator parameters become 26.
All features (continuous, binary, and one-hot) are scaled to $[-1, 1]$.
Global features are broadcast to all six operators and concatenated with per-operator features, producing a 47-dimensional feature vector per node.

\subsubsection{FM-Inspired Message Passing}
\label{sec:message_passing}

We design a message passing layer inspired by FM synthesis: messages correspond to modulation signals, aggregation combines incoming modulation, and FiLM-based updates condition each operator's processing on the aggregated modulation.

\textbf{Node features.}
An input MLP projects node $v_i$'s 47-dim features to \textbf{base features} $h_i \in \mathbb{R}^{H}$ ($H{=}384$), which remain fixed throughout message passing.
Each node also maintains a \textbf{message passing state} $s_i^{(\ell)} \in \mathbb{R}^{H}$ with $s_i^{(0)} = \mathbf{0}$, updated at each layer $\ell = 1, \dots, L$ ($L{=}9$).
This two-track design mirrors FM synthesis: operator configurations are fixed while modulation signals evolve across the network.

\textbf{Message computation and aggregation.}
At layer $\ell$, each node sends its previous state $s_j^{(\ell-1)}$ (already gated by output level, so silent operators send zero) along outgoing edges, scaled by a per-edge weight $w_{(j,i)}$.
At each destination $v_i$, incoming messages are summed over the in-neighbors $\mathcal{N}(i)$, the nodes with a directed edge into $v_i$:
\begin{equation}
m_i^{(\ell)} = \sum_{j \in \mathcal{N}(i)} w_{(j,i)} \cdot s_j^{(\ell-1)}
\end{equation}
Modulation edges carry weight $w_{(j,i)} = 1$, passing the source state through unchanged.
Feedback edges carry weight $f_{\text{remap}}(p_{\text{fb}}) \cdot f_b$, where $p_{\text{fb}} \in [0,1]$ is the preset's global feedback parameter, $f_{\text{remap}}$ is a learned monotonic remapping, and $f_b$ is a learnable scalar initialized to 0.5.\footnote{The 0.5 initialization is derived from the Dexed source code.
The original DX7 maps the feedback parameter to a gain that varies by algorithm.
For rendering audio, we normalize this so all 32 algorithms share the same feedback-to-gain mapping.}

The remapping $f_{\text{remap}}$ is a piecewise-linear function on a uniform grid over $[0, 1]$.
Each segment between grid points has a learnable slope, stored in log-space so that after exponentiation every slope is positive, which guarantees monotonicity.
The slopes are normalized to sum to 1, so the function maps $[0,1]$ onto $[0,1]$ and is the identity at initialization.
The feedback remapper uses a grid of 10 points, enough to interpolate the DX7's eight discrete feedback levels.

\textbf{FiLM-conditioned update.}
Feature-wise Linear Modulation (FiLM)~\cite{perez_film_2017} conditions the processing of base features on the aggregated modulation $m_i^{(\ell)}$.
In GNN-FiLM~\cite{brockschmidt_gnn-film_2020}, target node representations modulate incoming messages.
We invert this: aggregated modulation signals modulate the processing of \emph{fixed} base features, mirroring how modulator outputs control carrier behavior in FM synthesis.
Each layer $k = 1, \dots, K$ ($K{=}5$) of a FiLM MLP has its own pair of linear maps: one generates a scale $\gamma_k$ and shift $\beta_k$ from the aggregated messages,
\begin{equation}
\gamma_k,\, \beta_k = \text{split} \!\Big(\text{Linear}\big(m_i^{(\ell)}\big)\Big),
\end{equation}
and the other transforms the previous layer's output $u_{k-1}$, starting from the base features $u_0 = h_i$:
\begin{equation}
u_k = \text{ReLU}\Big(\text{Linear}(u_{k-1}) \cdot (1 + \gamma_k) + \beta_k\Big) .
\end{equation}
Each hidden layer thus applies: Linear $\to$ FiLM $\to$ Dropout (0.3) $\to$ ReLU.
The FiLM MLP keeps every $u_k$ at $H$ dimensions, and its final layer omits dropout and activation.
We write the full MLP's output as $u_K = \text{FiLM}(h_i,\, m_i^{(\ell)})$.

FiLM layers are initialized at $\frac{1}{10}\times$ the default LeCun normal variance, so $\gamma \approx 0$ and $\beta \approx 0$ and FiLM is near-identity at the start of training.
This lets the base feature transformation train before message-dependent modulation emerges.

\textbf{Output level gating.}
\label{sec:ol_gating}
In real FM synthesis, an operator with Output Level $= 0$ produces silence regardless of its other parameters.
Rather than require the network to learn this from data, we enforce it architecturally.
After the FiLM MLP, layer normalization and a residual connection from $h_i$, the result is multiplied by the operator's remapped output level $\hat{o}_i = f_{\text{OL}}(o_i)$:
\begin{equation}
s_i^{(\ell)} = \Big(\text{LayerNorm}\!\big(\text{FiLM}(h_i,\, m_i^{(\ell)})\big) + h_i\Big) \cdot \hat{o}_i
\end{equation}
The output level remapper uses the same piecewise-linear architecture as the feedback remapper but with a 40-point grid, needed for the output level curve's more complex shape.
Since $s_j^{(\ell)}$ is already scaled by $\hat{o}_j$, messages from silent operators ($o_j = 0$) are exactly zero at layer $\ell + 1$, propagating the zero-signal constraint through the network.

\subsubsection{Carrier Aggregation and Output Projection}

After $L$ message passing layers, we sum the carrier representations, mirroring how carrier signals are summed in FM synthesis.
We use binary masks $M_{\text{alg}} \in \{0,1\}^{6}$ derived from the algorithm definition (not learned).
The carrier mask is known for held out algorithms because the topology is always provided as input:
\begin{equation}
h_{\text{carrier}} = \sum_{i=1}^{6} M_{\text{alg}}[i] \cdot s_i^{(L)}
\end{equation}
An output MLP maps $h_{\text{carrier}}$ to the DX7-GNN latent ($y_P \in \mathbb{R}^{512}$), which feeds into SLAP's projector and predictor networks, producing $z_P, q_P \in \mathbb{R}^{d}$.

\subsubsection{Topology Generalization via Shared Weights}
\label{sec:shared_weights}
All operator nodes share the same message passing weights across all $L$ layers, so the encoder can process any FM topology without retraining, given enough layers for the requested graph's diameter.
Extra depth is also harmless in our design.
In many GNNs, each layer updates a node's state from its own previous state.
After many layers, the node representations drift toward one another until they become indistinguishable, a failure known as oversmoothing~\cite{rusch2023surveyoversmoothinggraphneural}.
Our update instead recomputes each state from the fixed base features $h_i$ and the current messages $m_i^{(\ell)}$, never from the node's own previous state $s_i^{(\ell)}$.
On feedforward paths, once $L$ exceeds the graph diameter, the messages stop changing, and additional layers return the same output.
Feedback edges create cycles whose states need not converge, but our experiments show that additional depth does not degrade retrieval.

\section{Experimental Setup}
\label{sec:experiments}

\subsection{Dataset}

We use DX7AllTheWeb,\footnote{\url{https://bobbyblues.recup.ch/yamaha_dx7/dx7_patches.html}} a collection of DX7 presets previously used in research~\cite{shin_synthrl_2025,combes_neural_2025,le_vaillant_latent_2024,caspe_ddx7_2022,turian_one_2021,vaillant_improving_2021,barkan_inversynth_2023}.
After removing duplicates by exact parameter values (ignoring voice names), we filter to 31,443 presets whose 2-second renders (C4, velocity 85) exceed \qty{-40}{\text{LUFS}}.
Some duplicate or near-duplicate audio remains.
DX7 symmetries, such as interchangeable operators with swapped settings, let distinct presets render identical audio.

For training, audio is rendered on-the-fly as 4-second clips (\qty{3}{\s} note, \qty{1}{\s} release, MIDI pitch C4, velocity 85) at \qty{44.1}{\kilo\hertz}, downsampled to \qty{22.05}{\kilo\hertz}, using Python bindings\footnote{\url{https://github.com/DBraun/dexed-py}} to the Dexed\footnote{\url{https://github.com/asb2m10/dexed}} DX7 synthesizer.
CPU rendering on 40 cores is the training bottleneck, not GPU compute.

\begin{figure}[t]
\centering
\includegraphics[width=\columnwidth]{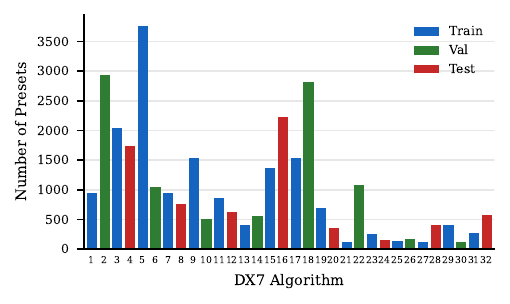}
\caption{Distribution of 31,443 presets across the 32 DX7 algorithms, colored by the held out split: train (odd algorithms);
validation and test (disjoint even algorithms).
The count per algorithm ranges from 116 (Algorithm 21) to 3,758 (Algorithm 5).}
\label{fig:algorithm_distribution}
\end{figure}

\textbf{Data splits.}
To test topology generalization, we use an interleaved algorithm split:
train on 16 odd-numbered algorithms (1, 3, \ldots, 31; 15,370 presets), validate on 8 even algorithms (2, 6, 10, \ldots, 30; 9,233 presets), and test on the remaining 8 even algorithms (4, 8, 12, \ldots, 32; 6,840 presets), from which a fixed random subset of 4,096 presets forms the retrieval gallery.
Before the random subset, Figure~\ref{fig:algorithm_distribution} shows the imbalanced distribution across algorithms, colored by the held out split.
The interleaved design balances structural properties across splits since higher numbered algorithms tend to have fewer modulators and more carriers.
To evaluate retrieval when every topology is seen during training, we also use an 80/10/10 random split across all 32 algorithms.

\textbf{Data augmentation.}
For training only, we augment parameters in two ways.
First, operator swapping: for every training sample, each non-silent operator (output level $> 0.05$) is independently replaced with 20\% probability.
Replacements are drawn from a global pool of all non-silent operator settings in the training set, maintaining the carrier/modulator distinction (carriers are only replaced by carriers, modulators by modulators).
All parameters of the operator are replaced together.
Second, parameter noise: each continuous parameter is independently interpolated toward a uniform random value with 1\% probability, blending original and random values by a random mixing coefficient $\alpha \in [0, 1]$.
Categorical parameters (LFO waveform, breakpoint curves) are perturbed to adjacent values with 2\% probability.

\subsection{Model Architectures}

For the audio encoder, we use AFx-Rep's~\cite{steinmetz_st-ito_2024} architecture variation of CNN14 from PANNs~\cite{kong_panns_2020}, trained from scratch with a 768-dimensional latent $y_A$.
We use SynthRL's mel-spectrogram settings~\cite{shin_synthrl_2025}.

\textbf{Transformer.}
\label{sec:transformer_baseline}
Our Transformer~\cite{vaswani2017attention} baseline treats the six operators the way a language model treats words in a sentence.
Each operator becomes a token, and self-attention lets every operator attend to every other.
The tokens come from the same input pipeline as the DX7-GNN (global features broadcast to all six operators, concatenated with per-operator features, then projected to hidden dimension $H$ by an input MLP), with learnable positional embeddings and a CLS token for pooling.
The Transformer keeps output level and feedback in its node features (49 dims vs.\ the DX7-GNN's 47), since it lacks the gating and feedback edge that handle them architecturally.

While the DX7-GNN reads the topology from its edge structure, the Transformer only knows the algorithm categorically.
Each Transformer block applies self-attention over the operator tokens, then cross-attention to a projection of the global features and a learnable per-algorithm embedding, then a feed-forward network, with layer normalization before each sublayer (pre-norm).
An unseen algorithm has no trained embedding, so we apply algorithm dropout.
During training, the algorithm embedding is replaced with a learned ``unknown'' token with 10\% probability, and held out algorithms use this unknown token at test time.

\textbf{Highway Network.}
\label{sec:highway_baseline}
We adapt the HN-OH preset encoder from Synth-Proxy~\cite{combes_neural_2025} as a flat-vector baseline.
It uses the same one-hot feature extraction as the other encoders, retaining output level and feedback, and flattens the $6 \times 49$ node features into a single 294-dimensional vector, concatenated with a learned algorithm embedding.
Highway Network blocks~\cite{srivastava_training_2015} then process this vector.
Like the Transformer, it applies algorithm dropout (10\%) and uses the unknown token for held out evaluation.

\subsection{Implementation Details}

For the DX7-GNN, the GELU-activated input MLP expands features before projection (input ratios $[4]$: $47 \to 4H \to H$).
The output MLP similarly expands before the final projection (output ratios $[4]$: $H \to 4H \to 512$).
The Transformer uses 4 blocks with $H{=}1024$, 8 attention heads, FFN ratio 2.0, ReLU activation, and CLS-token pooling.
Its input MLP uses ratios $[4, 1]$: $49 \to 4H \to H \to H$.
The Highway encoder uses 6 blocks with $H{=}768$, batch normalization, and ReLU activation, matching the HN-OH configuration from~\cite{combes_neural_2025}.
Table~\ref{tab:param_counts} reports parameter counts.

\begin{table}[th]
\centering
\caption{Parameter counts for the three parameter encoders.
All models share the same audio encoder (82.8M) and SLAP projector/predictor heads (3.4--3.6M per arm, depending on encoder output dimension).}
\label{tab:param_counts}
\begin{tabular}{lrcc}
\toprule
Component & DX7-GNN & Transformer & Highway \\
\midrule
Input / Embedding & 667K & 6,568K & 3K \\
Blocks / Layers & 1,331K & 50,401K & 6,191K \\
Output / Projection & 1,379K & 264K & 394K \\
\midrule
\textbf{Total} & \textbf{3.38M} & \textbf{57.2M} & \textbf{6.59M} \\
\bottomrule
\end{tabular}
\end{table}

\textbf{Hyperparameter search.}
We select DX7-GNN architectural hyperparameters via a 100-trial search that uses precomputed SynthRL encoder latents~\cite{shin_synthrl_2025} as a frozen audio proxy ($\sim$524K augmented presets, pooled through a trainable attention layer).
Each trial trains only the parameter encoder and SLAP heads, avoiding on-the-fly rendering and end-to-end training.
The best settings not specific to the DX7-GNN (e.g., learning rate, SLAP settings) carry over to hyperparameter searches of the two baselines.
Finally, the best configurations are trained end-to-end with PANNs from scratch.

\textbf{Training procedure.}
We train on four Nvidia L40 GPUs with AdamW~\cite{loshchilov_decoupled_2019} ($\beta_1{=}0.9$, $\beta_2{=}0.999$, effective batch size 256, weight decay $10^{-4}$, gradient clipping at 3.0, learning rate $2 \times 10^{-4}$ with 1000-step linear warmup).
A constant learning rate slightly outperformed cosine decay.
Training the DX7-GNN for 80K steps takes under five hours.
In that time, the renderer synthesizes roughly 2.6 years of audio ($4\,\text{s} \times 256 \times \text{80K}$ steps), about \qty{14.5}{\tera\byte} of uncompressed mono float32 at \qty{44.1}{\kilo\hertz}.

\section{Results}
\label{sec:results}

We evaluate audio-to-parameter retrieval using Recall@K, Mean Reciprocal Rank (MRR), and modality gap ($L_2$ distance between $q_A$ and $q_P$ centroids)~\cite{guinot_slap_2025}.
Queries are the audio renders of the gallery presets, so each query has exactly one correct preset in the gallery.
Although possible to retrieve via $d_{\cos}(z_A, z_P)$, we follow SLAP and use $d_{\cos}(q_A, q_P)$.
Like~\cite{hayes_audio_2025}, we argue that metrics based on parameter domain distance would be misleading due to symmetries in the DX7 topologies.

\subsection{Training on All Algorithms}

Table~\ref{tab:80_10_10_split} compares all three parameter encoders where every topology appears in training.
All three achieve strong retrieval over a held out gallery of 3,144 presets.
The DX7-GNN and Transformer both obtain 86.1\% R@1 (a near-exact tie, separated by one query and $4 \times 10^{-4}$ MRR), and the Highway encoder 81.2\%.
When all topologies are seen, the choice of parameter encoder matters relatively little.
This makes performance on held out algorithms the key test of structural inductive bias and generalization.

\begin{table}[h]
\centering
\caption{
Architectural comparison with 80/10/10 random split across all 32 algorithms (3,144-preset gallery).
Bold marks the best value per column.
}
\label{tab:80_10_10_split}
{
\begin{tabular}{l|ccc|c}
\toprule
& \multicolumn{3}{c|}{Audio $\rightarrow$ Param} & Mod. Gap \\
Model & R@1 & R@10 & MRR & ($L_2$ Dist.) \\
\midrule
DX7-GNN & \textbf{86.1\%} & \textbf{99.5\%} & \textbf{0.917} & 0.0293 \\
Transformer & \textbf{86.1\%} & \textbf{99.5\%} & \textbf{0.917} & \textbf{0.0209} \\
Highway & 81.2\% & 98.5\% & 0.880 & 0.0268 \\
\bottomrule
\end{tabular}
}
\end{table}

\subsection{Training with Held Out Algorithms}
\label{sec:held_out}

The three encoders that performed within 5~pp of each other on seen topologies now diverge sharply with held out algorithms (Table~\ref{tab:held_out_algorithms}).
The DX7-GNN achieves 52.2\% R@1 and 88.5\% R@10, outperforming the Transformer (34.6\% R@1) and Highway encoder (24.8\% R@1) despite having the fewest parameters (Table~\ref{tab:param_counts}).

\begin{table}[t]
\centering
\caption{
    Retrieval on held out algorithms (4,096-preset gallery).
    Bold marks the best value per column.
}
\label{tab:held_out_algorithms}
{
\begin{tabular}{l|ccc|c}
\toprule
& \multicolumn{3}{c|}{Audio $\rightarrow$ Param} & Mod. Gap \\
Model & R@1 & R@10 & MRR & ($L_2$ Dist.) \\
\midrule
DX7-GNN & \textbf{52.2\%} & 88.5\% & \textbf{0.652} & 0.0446 \\
Transformer & 34.6\% & 70.6\% & 0.470 & 0.0505 \\
Highway & 24.8\% & 59.1\% & 0.365 & 0.0722 \\
\midrule
\multicolumn{5}{l}{\textit{DX7-GNN augmentation ablation}} \\
No Swap & 13.5\% & 47.0\% & 0.245 & 0.0617 \\
No Noise & 50.5\% & 87.9\% & 0.640 & 0.0482 \\
\midrule
\multicolumn{5}{l}{\textit{DX7-GNN architecture ablation}} \\
No Feedback & 49.5\% & \textbf{89.2\%} & 0.636 & \textbf{0.0410} \\
Frozen Feedback & 51.2\% & 87.9\% & 0.644 & 0.0450 \\
Fully Connected & 30.9\% & 68.1\% & 0.436 & 0.0599 \\
\bottomrule
\end{tabular}
}
\end{table}

\textbf{Operator swapping is essential.}
Without it, test R@1 collapses from 52.2\% to 13.5\%.
Swapping exposes the model to novel operator settings within each topology.
By contrast, removing parameter noise only reduces R@1 to 50.5\%.

\textbf{Feedback helps, through message passing.}
The DX7-GNN receives the feedback level only through the message passing edge weight, so any benefit from feedback must arise there.
Freezing the feedback edge weight at its DX7-derived $0.5\times$ initialization, which removes the learned remapping, reduces R@1 only slightly, from 52.2\% to 51.2\%.
Removing feedback from the encoder entirely, by forcing the edge weight to zero while still rendering audio \emph{with} feedback, reduces R@1 to 49.5\%.
The encoder thus derives a small but consistent benefit at R@1 (learned 52.2\%, frozen 51.2\%, none 49.5\%).
However, the No Feedback ablation slightly surpasses the full model at R@10 (89.2\% vs.\ 88.5\%), so the benefit is limited to fine-grained top-1 retrieval.

\textbf{The specific routing matters.}
Replacing each preset's algorithm topology with a fully connected graph (all six operators mutually connected, ignoring the algorithm) collapses R@1 to 30.9\%.
The DX7-GNN's generalization thus stems from encoding the actual signal routing, not merely from message passing among the six operators.

\subsection{Message Passing Depth}

\begin{figure}[ht]
\centering
\includegraphics[width=\columnwidth]{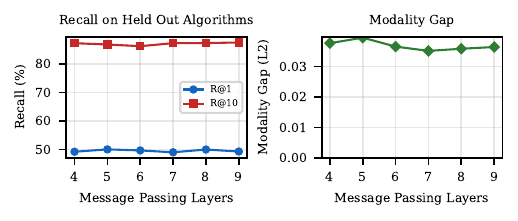}
\centering
\caption{Message passing depth ablation (4--9 layers).}
\label{fig:message_passing_ablation}
\end{figure}

With held out algorithms, Figure~\ref{fig:message_passing_ablation} varies message passing depth from four to nine layers, holding all else fixed and training each model at a quarter batch size.
Retrieval and modality gap are nearly flat across depths, supporting the DX7-GNN's shared-weight design.
Ignoring feedback cycles, just four layers already cover the largest graph diameter of Algorithms 1 and 2.
Extra layers neither improve retrieval much nor degrade it through oversmoothing.

\subsection{Learned Monotonic Remappings}
\label{sec:remappings}

\begin{figure}[t]
\centering
\includegraphics[width=\columnwidth]{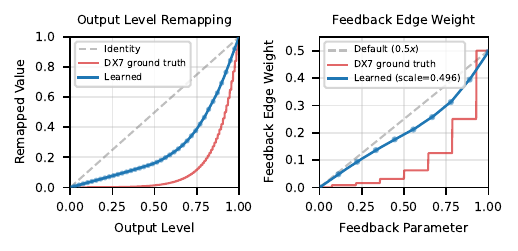}
\caption{Learned monotonic remappings overlaid with DX7 ground truth.
(Left) Output level: the learned curve tracks the DX7's exponential gain response.
(Right) Feedback edge weight, scaled by the 0.5 factor from Section~\ref{sec:message_passing}: the learned curve suppresses mid-range values, moving toward the DX7's power-of-two staircase.
}
\label{fig:remappings}
\end{figure}

Figure~\ref{fig:remappings} shows the two learned monotonic remappings overlaid with the DX7's ground truth response curves, derived from the Dexed source code.
The output level remapper learns a convex function similar to the DX7's exponential gain curve, guided by the SLAP objective alone, without access to internal gain tables.
For the feedback edge weight, the remapper $f_{\text{remap}}$ learns a mild suppression of mid-range values, moving toward the DX7's power-of-two staircase while remaining smooth, and the strength scalar $f_b$ converges to $0.496$, essentially unchanged from its $0.5$ initialization.

Our architecture induces a linearity property.
Node states combine by addition, as signals do in the DX7, and the learned remappings rescale features to suit this additive representation space.
This property may be similar to the linearity that Torres et al.~\cite{torres_learning_2026} induce in a neural audio codec via data augmentation.

\subsection{Timbre Understanding}

To test whether the audio encoder captures perceptually meaningful structure beyond DX7 retrieval, we evaluate on the timbremetrics benchmark~\cite{tian_assessing_2025}, which measures triplet agreement between embedding distances and human timbre similarity ratings across 334 audio samples.
Despite training exclusively on FM synthesis, all three audio encoders achieve roughly 68--71\% triplet agreement (Table~\ref{tab:timbremetrics}), comparable to LAION-CLAP~\cite{wu_large-scale_2023} (70.8--71.8\%), the strong MFCC baseline (68.7\%), and well above chance (50\%).
A similar but slightly more expressive CNN14 backbone with off-the-shelf AudioSet weights (PANNs~\cite{kong_panns_2020}) reaches only 61.4\%, below MFCC.
This is consistent with the benchmark's finding that MFCC outperforms many trained audio models~\cite{tian_assessing_2025}.

\begin{table}[t]
\centering
\caption{
    Triplet agreement (\%) on timbremetrics~\cite{tian_assessing_2025}.
    $y_A$: encoder output; $z_A$: projection; $q_A$: prediction.
    Bold marks the best value per column.
}
\label{tab:timbremetrics}
{
\begin{tabular}{lccc}
\toprule
Model & $y_A$ & $z_A$ & $q_A$ \\
\midrule
DX7-GNN & 68.4\% & 69.3\% & 69.3\% \\
Transformer & 71.0\% & \textbf{71.3\%} & \textbf{71.1\%} \\
Highway & 68.3\% & 68.7\% & 68.1\% \\
\midrule
MFCC & 68.7\% & --- & --- \\
LAION-CLAP & \textbf{71.8\%} & 70.8\% & --- \\
PANNs (AudioSet) & 61.4\% & --- & --- \\
\bottomrule
\end{tabular}
}
\end{table}

\subsection{Qualitative Results}

\begin{figure}[t]
\centering
\includegraphics[width=\columnwidth]{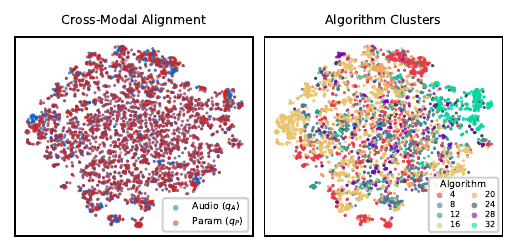}
\caption{
    t-SNE visualization of held out algorithm embeddings.
    (Left) Audio and parameter embeddings by color.
    (Right) Embeddings from both modalities colored by algorithm.
}
\label{fig:tsne}
\end{figure}

Figure~\ref{fig:tsne} visualizes the learned embedding space via t-SNE.
Audio and parameter embeddings form well-aligned clusters with minimal modality gap, confirming that SLAP training produces a joint representation~\cite{guinot_slap_2025}.
Presets from the same algorithm tend to cluster together, though overlap between algorithms suggests that different topologies can produce perceptually similar sounds.
Retrieval quality has not been assessed in a user study, though our interactive website offers a sandbox for subjective evaluation.

\section{Conclusion}

We presented DX7-GNN and FM-SynAPSE for joint embeddings of audio and FM synthesizer parameters.
A controlled comparison with Transformer and Highway Network encoders shows that the graph-structure inductive bias improves generalization to new topologies.
Operator swapping augmentation is essential for this generalization.
Held out performance is stable across message passing depths from four to nine layers, consistent with our hypothesis of convergence without oversmoothing.

\textbf{Limitations.}
Our evaluation covers only the DX7, whose structurally identical operators provide a convenient setting for studying topological generalization.
The 31K presets are imbalanced across algorithms, and every preset is rendered with a single note (C4, velocity 85), so the embeddings capture one point in each preset's pitch- and velocity-dependent behavior.
All results represent single training runs, and architectural choices were validated through hyperparameter search with frozen audio encoder latents rather than a costlier end-to-end search.

\textbf{Future directions.}
Next steps include moving beyond retrieval to optimization or generation.
A parameter encoder enables optimization in embedding space without rendering audio at every step~\cite{cherep_creative_2024}.
The DX7-GNN could support matching sounds with fewer operators and exploration of FM configurations beyond the DX7's 32 algorithms.
Exploration of the audio encoder's design space while training on polyphonic MIDI could yield a more useful model.
GNN shared-weight message passing could extend to effect chains, mixing graphs, or modular synthesis, though diverse module types (filters, envelopes, LFOs) may require heterogeneous nodes or edges.
One path is to derive the graph from the intermediate representation of a language for audio processing such as Faust~\cite{Orlarey2009}.

\section{Acknowledgments}
This work was supported in part by NSF Award 2523649.

\bibliographystyle{IEEEtranDAFx}
\bibliography{main}

@inproceedings{shin_synthrl_2025,
	title = {{SynthRL}: {Cross}-domain {Synthesizer} {Sound} {Matching} via {Reinforcement} {Learning}},
	booktitle = {Proceedings of the {Thirty}-{Fourth} {International} {Joint} {Conference} on {Artificial} {Intelligence}},
	series = {IJCAI '25},
	author = {Shin, Wonchul and Lee, Kyogu},
	year = {2025},
}

@incollection{Orlarey2009,
	Author = {Yann Orlarey and St\'ephane Letz and Dominique Fober},
    Title = {{Faust}: an {Efficient} {Functional} {Approach} to {DSP} {Programming}},
	Booktitle = {{New} {Computational} {Paradigms} for {Computer} {Music}},
	Year = {2009},
	pages = {65--96}
}

@inproceedings{wu_large-scale_2023,
	title = {Large-scale {Contrastive} {Language}-{Audio} {Pretraining} with {Feature} {Fusion} and {Keyword}-to-{Caption} {Augmentation}},
	doi = {10.1109/ICASSP49357.2023.10095969},
	booktitle = {{IEEE} {International} {Conference} on {Acoustics}, {Speech} and {Signal} {Processing} ({ICASSP})},
	author = {Wu, Yusong and Chen, Ke and Zhang, Tianyu and Hui, Yuchen and Nezhurina, Marianna and Berg-Kirkpatrick, Taylor and Dubnov, Shlomo},
	year = {2023},
}

@article{vaswani2017attention,
    title={{Attention} is {All} {You} {Need}},
    author={Vaswani, Ashish and Shazeer, Noam and Parmar, Niki and Uszkoreit, Jakob and Jones, Llion and Gomez, Aidan N and Kaiser, {\L}ukasz and Polosukhin, Illia},
    journal={Advances in Neural Information Processing Systems},
    volume={30},
    year={2017}
}

@inproceedings{
    loshchilov_decoupled_2019,
    title={{Decoupled} {Weight} {Decay} {Regularization}},
    author={Ilya Loshchilov and Frank Hutter},
    booktitle={International Conference on Learning Representations},
    year={2019},
}

@inproceedings{turian_one_2021,
    author={Turian, Joseph and Shier, Jordie and Tzanetakis, George and McNally, Kirk and Henry, Max},
    booktitle={2021 {24th} {International} {Conference} on {Digital} {Audio Effects} (DAFx)},
    title={{One} {Billion} {Audio} {Sounds} from {GPU}-{Enabled} {Modular} {Synthesis}},
    year={2021},
    volume={},
    number={},
    pages={222-229},
    doi={10.23919/DAFx51585.2021.9768246}
}

@article{yee-king_automatic_2018,
	title = {Automatic {Programming} of {VST} {Sound} {Synthesizers} {Using} {Deep} {Networks} and {Other} {Techniques}},
	volume = {2},
	number = {2},
	journal = {IEEE Transactions on Emerging Topics in Computational Intelligence},
	author = {Yee-King, Matthew John and Fedden, Leon and d'Inverno, Mark},
	year = {2018},
	pages = {150--159},
}

@inproceedings{ye_nas-fm_2023,
    author = {Ye, Zhen and Xue, Wei and Tan, Xu and Liu, Qifeng and Guo, Yike},
    title = {{NAS-FM}: {Neural} {Architecture} {Search} for {Tunable} and {Interpretable} {Sound} {Synthesis} based on {Frequency} {Modulation}},
    year = {2023},
    isbn = {978-1-956792-03-4},
    doi = {10.24963/ijcai.2023/651},
    booktitle = {Proceedings of the Thirty-Second International Joint Conference on Artificial Intelligence},
    articleno = {651},
    numpages = {9},
    pages = {5869--5877},
    location = {Macao, P.R.China},
    series = {IJCAI '23}
}

@article{masuda_quality-diversity_2023,
	author="Masuda, Naotake and Saito, Daisuke",
	title="{Quality}-diversity for {Synthesizer} {Sound} {Matching}",
	journal="Journal of Information Processing",
	publisher="Information Processing Society of Japan",
	year="2023",
	volume="31",
	pages="220-228",
	DOI="10.2197/ipsjjip.31.220",
}

@inproceedings{caspe_ddx7_2022,
	booktitle = {Proceedings of the 23rd {International} {Society} for {Music} {Information} {Retrieval} {Conference}},
    title={{DDX7: Differentiable FM Synthesis of Musical Instrument Sounds}},
    author={Caspe, Franco and McPherson, Andrew and Sandler, Mark},
    year={2022}
}

@inproceedings{chen_sound2synth_2022,
	title = {{Sound2Synth}: {Interpreting} {Sound} via {FM} {Synthesizer} {Parameters} {Estimation}},
	author    = {Chen, Zui and Jing, Yansen and Yuan, Shengcheng and Xu, Yifei and Wu, Jian and Zhao, Hang},
	booktitle = {Proceedings of the Thirty-First International Joint Conference on
			   Artificial Intelligence},
	series    = {IJCAI '22},
	pages     = {4921--4928},
	year      = {2022},
}

@inproceedings{vaillant_improving_2021,
	title = {Improving {Synthesizer} {Programming} {From} {Variational} {Autoencoders} {Latent} {Space}},
	booktitle = {2021 24th {International} {Conference} on {Digital} {Audio} {Effects} ({DAFx})},
	author = {Le Vaillant, Gwendal and Dutoit, Thierry and Dekeyser, Sebastien},
	year = {2021},
	pages = {276--283},
}

@inproceedings{yang_white_2023,
	title = {White {Box} {Search} {Over} {Audio} {Synthesizer} {Parameters}},
	booktitle = {Proceedings of the 24th {International} {Society} for {Music} {Information} {Retrieval} {Conference}},
	author = {Yang, Yuting and Jin, Zeyu and Barnes, Connelly and Finkelstein, Adam},
	year = {2023},
}

@inproceedings{elizalde_clap_2023,
	title = {{CLAP} {Learning} {Audio} {Concepts} from {Natural} {Language} {Supervision}},
	booktitle = {{IEEE} {International} {Conference} on {Acoustics}, {Speech} and {Signal} {Processing} ({ICASSP})},
	author = {Elizalde, Benjamin and Deshmukh, Soham and Ismail, Mahmoud Al and Wang, Huaming},
	year = {2023},
}

@misc{uzrad_diffmoog_2024,
	title = {{DiffMoog}: a {Differentiable} {Modular} {Synthesizer} for {Sound} {Matching}},
	shorttitle = {{DiffMoog}},
	url = {http://arxiv.org/abs/2401.12570},
    eprint = {2401.12570},
    archivePrefix = {arXiv},
	author = {Uzrad, Noy and Barkan, Oren and Elharar, Almog and Shvartzman, Shlomi and Laufer, Moshe and Wolf, Lior and Koenigstein, Noam},
	year = {2024}
}

@inproceedings{brade_synthscribe_2024,
    author = {Brade, Stephen and Wang, Bryan and Sousa, Mauricio and Newsome, Gregory Lee and Oore, Sageev and Grossman, Tovi},
    title = {{SynthScribe}: {Deep} {Multimodal} {Tools} for {Synthesizer} {Sound} {Retrieval} and {Exploration}},
    year = {2024},
    isbn = {9798400705083},
    publisher = {Association for Computing Machinery},
    doi = {10.1145/3640543.3645158},
    booktitle = {Proceedings of the 29th International Conference on Intelligent User Interfaces},
    pages = {51–65},
    numpages = {15},
    location = {Greenville, SC, USA},
    series = {IUI '24}
}

@inproceedings{cherep_creative_2024,
  title = 	 {Creative {Text}-to-{Audio} {Generation} via {Synthesizer} {Programming}},
  author =       {Cherep, Manuel and Singh, Nikhil and Shand, Jessica},
  booktitle = 	 {Proceedings of the 41st International Conference on Machine Learning},
  pages = 	 {8270--8285},
  year = 	 {2024},
  volume = 	 {235},
  series = 	 {Proceedings of Machine Learning Research},
  publisher =    {PMLR},
}

@inproceedings{bruford_synthesizer_2024,
	title = {Synthesizer {Sound} {Matching} {Using} {Audio} {Spectrogram} {Transformers}},
    booktitle = {2024 {27th} {International} {Conference} on {Digital} {Audio Effects} (DAFx)},
	author = {Bruford, Fred and Blang, Frederik and Nercessian, Shahan},
	year = {2024},
	note = {{Late Breaking Results}},
}

@inproceedings{steinmetz_st-ito_2024,
	title = {{ST}-{ITO}: {Controlling} {Audio} {Effects} for {Style} {Transfer} with {Inference}-{Time} {Optimization}},
	booktitle = {Proceedings of the 25th {International} {Society} for {Music} {Information} {Retrieval} {Conference}},
	author = {Steinmetz, Christian J. and Singh, Shubhr and Comunità, Marco and Ibnyahya, Ilias and Yuan, Shanxin and Benetos, Emmanouil and Reiss, Joshua D.},
	year = {2024},
}

@mastersthesis{shier_synthesizer_2021,
	author = {Shier, Jordie},
	title = {The {Synthesizer} {Programming} {Problem}: {Improving} the {Usability} of {Sound} {Synthesizers}},
	school = {University of Victoria},
	year = {2021},
	address = {Victoria, BC, Canada},
}

@article{chowning_synthesis_1973,
	title = {The {Synthesis} of {Complex} {Audio} {Spectra} by {Means} of {Frequency} {Modulation}},
	volume = {21},
	number = {7},
	journal = {Journal of the Audio Engineering Society},
	author = {Chowning, John M.},
	year = {1973},
	pages = {526--534},
}

@article{kong_panns_2020,
    author = {Kong, Qiuqiang and Cao, Yin and Iqbal, Turab and Wang, Yuxuan and Wang, Wenwu and Plumbley, Mark D.},
    title = {{PANNs}: {Large}-{Scale} {Pretrained} {Audio} {Neural} {Networks} for {Audio} {Pattern} {Recognition}},
    year = {2020},
    issue_date = {2020},
    volume = {28},
    issn = {2329-9290},
    doi = {10.1109/TASLP.2020.3030497},
    journal = {IEEE/ACM Transactions on Audio, Speech, and Language Processing},
    pages = {2880–2894},
    numpages = {15}
}

@inproceedings{cherep_contrastive_2025,
    title={{Contrastive} {Learning} from {Synthetic} {Audio} {Doppelg\"angers}},
    author={Manuel Cherep and Nikhil Singh},
    booktitle={The Thirteenth International Conference on Learning Representations},
    year={2025},
}

@article{le_vaillant_latent_2024,
	title = {Latent {Space} {Interpolation} of {Synthesizer} {Parameters} {Using} {Timbre}-{Regularized} {Auto}-{Encoders}},
	volume = {32},
	journal = {IEEE/ACM Transactions on Audio, Speech, and Language Processing},
	author = {Le Vaillant, Gwendal and Dutoit, Thierry},
	year = {2024},
	pages = {3379--3392},
}

@inproceedings{barkan_inversynth_2023,
	title = {{InverSynth} {II}: {Sound} {Matching} {via} {Self}-{Supervised} {Synthesizer}-{Proxy} {and} {Inference}-{Time} {Finetuning}},
	booktitle = {Proceedings of the 24th {International} {Society} for {Music} {Information} {Retrieval} {Conference}},
	author = {Barkan, Oren and Shvartzman, Shlomi and Uzrad, Noy and Laufer, Moshe and Elharar, Almog and Koenigstein, Noam},
	year = {2023},
}

@inproceedings{guinot_slap_2025,
	title = {{SLAP}: {Siamese} {Language}-{Audio} {Pretraining} {Without} {Negative} {Samples} for {Music} {Understanding}},
	booktitle = {Proceedings of the 26th {International} {Society} for {Music} {Information} {Retrieval} {Conference}},
	author = {Guinot, Julien and Riou, Alain and Quinton, Elio and Fazekas, György},
	year = {2025},
}

@inproceedings{hayes_audio_2025,
	title = {Audio {Synthesizer} {Inversion} in {Symmetric} {Parameter} {Spaces} with {Approximately} {Equivariant} {Flow} {Matching}},
	booktitle = {Proceedings of the 26th {International} {Society} for {Music} {Information} {Retrieval} {Conference}},
	author = {Hayes, Ben and Saitis, Charalampos and Fazekas, György},
	year = {2025},
}

@inproceedings{tian_assessing_2025,
	title = {Assessing the {Alignment} of {Audio} {Representations} with {Timbre} {Similarity} {Ratings}},
	booktitle = {Proceedings of the 26th {International} {Society} for {Music} {Information} {Retrieval} {Conference}},
	author = {Tian, Haokun and Lattner, Stefan and Saitis, Charalampos},
	year = {2025},
}

@inproceedings{grill_bootstrap_2020,
	title = {Bootstrap {Your} {Own} {Latent} - {A} {New} {Approach} to {Self}-{Supervised} {Learning}},
	volume = {33},
	booktitle = {Advances in {Neural} {Information} {Processing} {Systems}},
	author = {Grill, Jean-Bastien and Strub, Florian and Altché, Florent and Tallec, Corentin and Richemond, Pierre and Buchatskaya, Elena and Doersch, Carl and Avila Pires, Bernardo and Guo, Zhaohan and Gheshlaghi Azar, Mohammad and Piot, Bilal and kavukcuoglu, koray and Munos, Remi and Valko, Michal},
	year = {2020},
}

@article{combes_neural_2025,
	title = {Neural {Proxies} for {Sound} {Synthesizers}: {Learning} {Perceptually} {Informed} {Preset} {Representations}},
	volume = {73},
	number = {9},
	journal = {Journal of the Audio Engineering Society},
	author = {Combes, Paolo and Weinzierl, Stefan and Obermayer, Klaus},
	year = {2025},
	pages = {561--577},
}

@inproceedings{vaillant_contrastive_2026,
	title = {Contrastive {Timbre} {Representations} for {Musical} {Instrument} and {Synthesizer} {Retrieval}},
	booktitle = {{IEEE} {International} {Conference} on {Acoustics}, {Speech} and {Signal} {Processing} ({ICASSP})},
	author = {Le Vaillant, Gwendal and Molle, Yannick},
	year = {2026},
}

@misc{rusch2023surveyoversmoothinggraphneural,
      title={{A} {Survey} on {Oversmoothing} in {Graph} {Neural} {Networks}},
      author={T. Konstantin Rusch and Michael M. Bronstein and Siddhartha Mishra},
      year={2023},
      eprint={2303.10993},
      archivePrefix={arXiv},
      url={https://arxiv.org/abs/2303.10993},
}

@inproceedings{brockschmidt_gnn-film_2020,
	title = {{GNN}-{FiLM}: {Graph} {Neural} {Networks} with {Feature}-wise {Linear} {Modulation}},
	booktitle = {Proceedings of the 37th {International} {Conference} on {Machine} {Learning}},
	publisher = {PMLR},
	author = {Brockschmidt, Marc},
	year = {2020},
	pages = {1144--1152},
}

@inproceedings{perez_film_2017,
    author = {Perez, Ethan and Strub, Florian and de Vries, Harm and Dumoulin, Vincent and Courville, Aaron},
    title = {{FiLM}: {Visual} {Reasoning} with a {General} {Conditioning} {Layer}},
    year = {2018},
    isbn = {978-1-57735-800-8},
	booktitle = {{Proceedings} of the {Thirty}-{Second} {AAAI} {Conference} on {Artificial} {Intelligence} ({AAAI}-18)},
	articleno = {483},
    numpages = {10},
}

@inproceedings{torres_learning_2026,
	title = {Learning {Linearity} in {Audio} {Consistency} {Autoencoders} via {Implicit} {Regularization}},
	booktitle = {{IEEE} {International} {Conference} on {Acoustics}, {Speech} and {Signal} {Processing} ({ICASSP})},
	author = {Torres, Bernardo and Moussallam, Manuel and Meseguer-Brocal, Gabriel},
	year = {2026},
}

@inproceedings{srivastava_training_2015,
	title = {Training {Very} {Deep} {Networks}},
	volume = {28},
	urldate = {2026-03-28},
	booktitle = {Advances in {Neural} {Information} {Processing} {Systems}},
	author = {Srivastava, Rupesh K and Greff, Klaus and Schmidhuber, Jürgen},
	year = {2015}
}

\end{document}